\documentclass[%
aps, prl,
floatfix,
preprintnumbers,
reprint,
superscriptaddress,
amsmath, amssymb,
longbibliography,
]{revtex4-1}

\usepackage[utf8]{inputenc}
\usepackage{graphicx}
\usepackage{bm}
\usepackage[breaklinks=true]{hyperref}
\hypersetup{bookmarksnumbered, pdfpagemode=UseOutlines, 
	pdfauthor={K.\ S.\ Das}, 
	pdftitle={Spin Injection and Detection via the Spin Anomalous Hall Effect in a Ferromagnetic Metal}, 
	pdfdisplaydoctitle, 
	colorlinks=true, citecolor=blue, filecolor=blue, linkcolor=blue, urlcolor=blue}

\begin{document}
	
	\title{Spin Injection and Detection via the Anomalous Spin Hall Effect in a Ferromagnetic Metal}
	
	\author{K.\ S.\ \surname{Das}}
	\email[e-mail: ]{k.s.das@rug.nl}
	\affiliation{University of Groningen, Zernike Institute for Advanced Materials, NL-9747 AG Groningen, The Netherlands}
	\author{W.\ Y.\ \surname{Schoemaker}}
	\affiliation{University of Groningen, Zernike Institute for Advanced Materials, NL-9747 AG Groningen, The Netherlands}
	\author{B.\ J.\ \surname{van Wees}}
	\email[e-mail: ]{b.j.van.wees@rug.nl}
	\affiliation{University of Groningen, Zernike Institute for Advanced Materials, NL-9747 AG Groningen, The Netherlands}
	\author{I.\ J.\ Vera-Marun}
	\email[e-mail: ]{ivan.veramarun@manchester.ac.uk}
	\affiliation{School of Physics and Astronomy, University of Manchester, Manchester M13 9PL, United Kingdom}
	
	\date{\today}
	
	\begin{abstract}
		We report a novel spin injection and detection mechanism via the anomalous Hall effect in a ferromagnetic metal. The anomalous spin Hall effect (ASHE) refers to the transverse spin current generated within the ferromagnet. We utilize the ASHE and its reciprocal effect to electrically inject and detect magnons in a magnetic insulator in a non-local geometry. Our experiments reveal that permalloy can have a higher spin injection and detection efficiency to that of platinum, owing to the ASHE. We also demonstrate the tunability of the ASHE via the orientation of the permalloy magnetization, thus creating new possibilities for spintronic applications.
	\end{abstract}
	
	
	
	\maketitle
	
	
	
	In non-magnetic metals with high spin-orbit coupling, a charge current generates a transverse spin current via the spin Hall effect (SHE) \cite{kimura_room-temperature_2007,sinova_spin_2015}. This type of spin current generation perpendicular to a charge current has a significant technological relevance for spin transfer torque devices \cite{liu_spin-torque_2011,liu_spin-torque_2012} and also for the electrical injection of magnons (quantized spin waves) in magnetic insulators \cite{kajiwara_transmission_2010,cornelissen_long-distance_2015,goennenwein_non-local_2015}.
	 The electrical injection and detection of magnons offer a distinct technological advantage for the integration of magnon spintronics into solid state devices, over other magnon generation mechanisms such as spin pumping by radiofrequency fields \cite{chumak_direct_2012} or the spin Seebeck effect due to a temperature gradient \cite{uchida_spin_2010}. In this regard Platinum (Pt), a normal metal with a large spin-orbit coupling, is the most commonly used material for the electrical generation (and detection) of magnons via SHE. Recent studies showed that ferromagnets can also be utilized for electrical detection of magnons via the inverse spin Hall effect (ISHE) \cite{miao_inverse_2013, tsukahara_self-induced_2014, seki_observation_2015, tian_manipulation_2016}. In particular, Tian \textit{et. al.} \cite{tian_manipulation_2016} reported that ISHE in a ferromagnetic cobalt was independent of its magnetization direction. 

		\begin{figure*}[tbp]
			\includegraphics*[angle=0, trim=0mm 0mm 0mm 0mm, width=170mm]{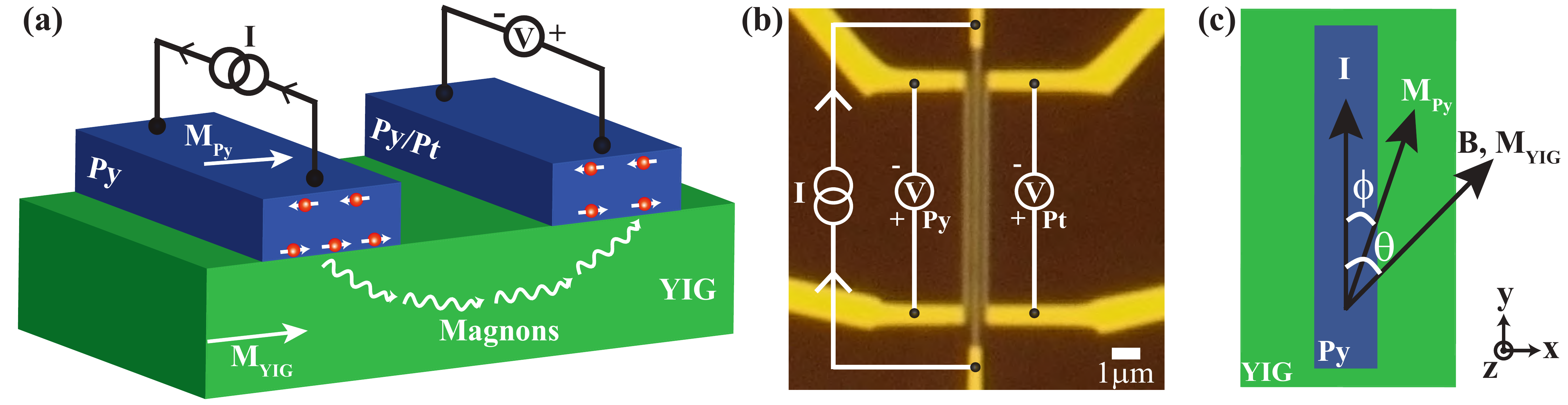}
			\caption{
				\label{fig:DeviceGeometry}
				\textbf{(a)} Schematic diagram of the experimental geometry. A charge current ($I$) through the Py injector generates a transverse spin accumulation at the Py/YIG interface via the ASHE and SHE, which excites magnons in YIG by the transfer of angular momentum. The reciprocal processes generate a non-local electrical voltage ($V$) at the detector. \textbf{(b)} Optical image of the device along with the illustration of the electrical connections. An alternating current ($I$) is sourced across the middle Py (injector) strip and the non-local voltages ($V_\text{Py}$ and $V_\text{Pt}$), generated across the left Py (detector) strip and the reference Pt (detector) strip on the right, are simultaneously measured. \textbf{(c)} An external in-plane magnetic field ($B$) is applied at an angle ($\theta$) with respect to the direction of $I$. The coercive field of our YIG film being very small ($\approx 1$~mT), the YIG magnetization ($M_\text{YIG}$) is parallel to $B$, while the Py magnetization ($M_\text{Py}$) makes an angle ($\phi$) with respect to $I$. 
			}
		\end{figure*}
		
	In a ferromagnetic metal the presence of the magnetization order parameter leads to the anomalous Hall effect (AHE) \cite{nagaosa_anomalous_2010}. Here, we report a novel mechanism of spin current generation in a ferromagnet related to the AHE. The AHE generates a transverse electric potential, mutually orthogonal to the applied charge current ($I$) in a FM and its magnetization ($M$) direction. Due to a finite spin polarization in a FM, we expect that AHE can also result in a transverse spin accumulation. We call this effect the anomalous spin Hall effect (ASHE) in a ferromagnet. In addition to this new ASHE, the regular SHE due to the spin-orbit coupling in the ferromagnetic material will also be present and contribute to a spin accumulation perpendicular to $I$. The spin accumulation due to SHE in the FM will be independent of $M$, since the inverse process (ISHE) in a FM was shown to be independent of its magnetization by Tian \textit{et. al.} \cite{tian_manipulation_2016}. To demonstrate this mechanism we realize for the first time non-local magnon transport in a ferrimagnetic insulator, yttrium iron garnet($\text{Y}_3\text{Fe}_5\text{O}_{12}$, YIG), with all-electrical injection and detection using a ferromagnetic metal, permalloy (Ni$_{80}$Fe$_{20}$, Py). The insulating spin transport channel (YIG) facilitates our observation of ASHE due to the lack of any parallel conducting path. Our experimental geometry is depicted in Fig.~\ref{fig:DeviceGeometry}(a). A charge current ($I$) sourced through a Py strip will result in a transverse spin accumulation. Given the presence of both a large spin-orbit coupling and a magnetization order parameter, we consider two contributions to the spin accumulation at the Py/YIG interface: i$)$ SHE, which is independent of the Py magnetization ($M_\text{Py}$) \cite{tian_manipulation_2016} and ii$)$ ASHE, which is maximized when $M_\text{Py}$ is perpendicular to the direction of $I$. This spin accumulation at the Py/YIG interface will generate magnons in the YIG by the transfer of angular momentum across the interface. Following the non-local magnon transport and its conversion into a pure spin current at the Py detector, there are reciprocal processes (ISHE and a magnetization-dependent inverse ASHE) that will generate an electrical voltage ($V$). Using a reference Pt detector, we directly compare the detection efficiencies of Py and Pt. Our experiments reveal that the detection efficiency of Py can exceed that of Pt when the contribution due to ASHE in the Py is tuned to its maximum value. 

The 210~nm thick YIG film used in this study is grown on GGG ($\text{Gd}_3\text{Ga}_5\text{O}_{12}$) substrate by liquid-phase epitaxy and obtained commercially from Matesy GmbH. Electron beam lithography was used to pattern the devices, which consist of two Py strips and one reference Pt strip, as shown in the optical image in Fig.~\ref{fig:DeviceGeometry}(b). The Py and Pt strips were deposited by d.c.\ sputtering in $\text{Ar}^+$ plasma. The Ti/Au leads and bonding pads were deposited by e-beam evaporation. The thicknesses of the Py and the Pt strips are 13~nm and 7~nm respectively, with widths of 200~nm. The electrical conductivities of the Py and Pt strips were measured to be $1.64\times10^6$~S/m and $4.71\times10^6$~S/m, respectively. The middle Py strip is used as the injector and the left Py strip and right Pt strip act as detectors. Both the Py and Pt detectors have the same geometry and are located 500~nm (centre-to-centre) away from the middle Py injector. The electrical connections for the non-local magnon transport experiment are shown schematically in Fig.~\ref{fig:DeviceGeometry}(b). An alternating current, with an amplitude of 350~$\mu$A and frequency of 11~Hz, is applied to the middle Py strip (injector). The non-local voltage across the left Py detector ($V_\text{Py}$) and across the reference Pt detector ($V_\text{Pt}$) are simultaneously recorded by a phase-sensitive lock-in detection technique. The linear signal corresponding to the electrical injection and detection is measured as the first harmonic (1$f$) response of the non-local voltage \cite{cornelissen_long-distance_2015}, while the thermally generated magnons due to Joule heating at the injector are detected as a Spin Seebeck signal, measured as the second harmonic (2$f$) response. For all our experiments, we normalize the detected non-local voltage ($V^\text{1(2)f}$) by the injection current ($I$) for the first harmonic response ($R_\text{NL}^\text{1f}=V^\text{1f}/I$) and by $I^2$ for the second harmonic response ($R_\text{NL}^\text{2f}=V^\text{2f}/I^2$). All measurements have been conducted under a low vacuum atmosphere at room temperature ($\approx 293$~K), using a superconducting magnet and a rotatable sample holder.
	 	
\begin{figure}[tbp]
	\includegraphics*[angle=0, trim=0mm 0mm 0mm 0mm, width=85mm]{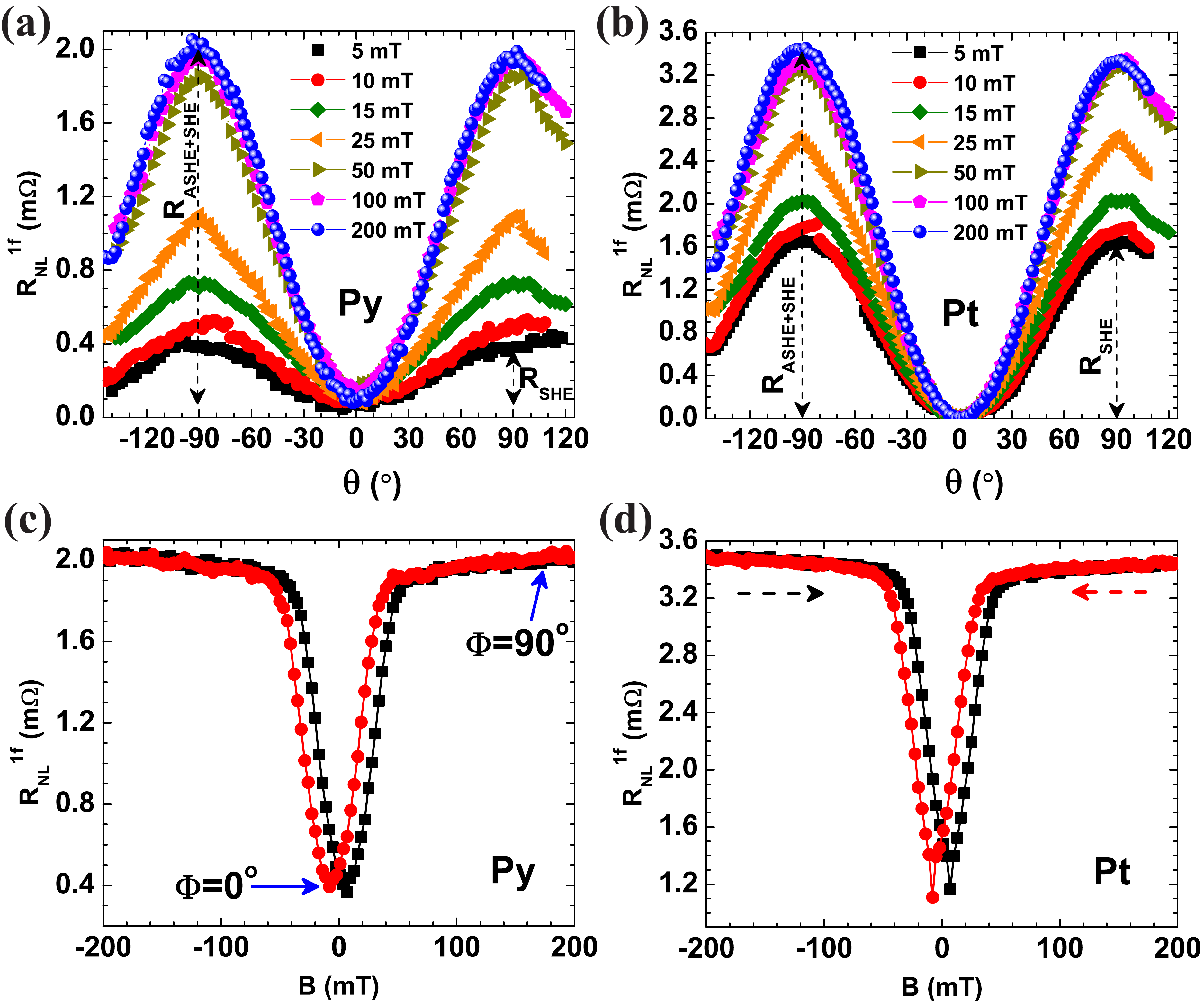}
	\caption{
		\label{fig:ExperimentalData}
		Non-local resistance ($R_\text{NL}^\text{1f}$) as a function of angle $\theta$ for different magnetic fields ($B$), measured by the Py detector \textbf{(a)} and by the reference Pt detector. \textbf{(b)}. Dependence of $R_\text{NL}^\text{1f}$ on $B$ at a fixed angle, $\theta = 90^\text{o}$, measured by the Py detector \textbf{(c)} and the Pt detector \textbf{(d)}. The black and the red curves represent trace and retrace of $B$ in the magnetic field sweep measurements, respectively. 
	}
\end{figure}	
	
	
	An external in-plane magnetic field ($B$) is applied at an angle $\theta$ with respect to the direction of the strips (and $I$), as shown in Fig.~\ref{fig:DeviceGeometry}(c). The coercive field of our YIG film is approximately 1~mT \cite{dejene_control_2015} and any $B$ greater than this value will cause the YIG magnetization ($M_\text{YIG}$) to align parallel to $B$. On the other hand, the Py strips have a shape anisotropy, which leads to a higher saturation field and to the Py magnetization ($M_\text{Py}$) fully aligning along $B$ only above 50~mT. In general, for $B<50$~mT, $M_\text{Py}$ makes an angle $\phi$ ($\neq \theta$) with respect to $I$. The experimental data is presented in Figs.~\ref{fig:ExperimentalData}(a-d). The non-local resistance, corresponding to the electrical generation and detection of the magnons, is measured as a function of the angle $\theta$ by the Py detector [$R_\text{NL}^\text{1f}$(Py)] and the Pt detector [$R_\text{NL}^\text{1f}$(Pt)], as shown in Figs.~\ref{fig:ExperimentalData}(a) and \ref{fig:ExperimentalData}(b), respectively. $R_\text{NL}^\text{1f}$(Py) and $R_\text{NL}^\text{1f}$(Pt) exhibit lineshapes resembling that of $\sin^2\theta$ \cite{cornelissen_long-distance_2015}. The angular dependence measurements are performed for different magnitudes of $B$. The amplitudes of both $R_\text{NL}^\text{1f}$(Py) and $R_\text{NL}^\text{1f}$(Pt) increase with $B$ and saturate above $B\approx50$~mT. This behaviour is confirmed in the $B$-sweep measurements at $\theta=90^\text{o}$, shown in Figs.~\ref{fig:ExperimentalData}(c) and \ref{fig:ExperimentalData}(d) for the Py and the Pt detectors, respectively.
	
	The $B$-dependence of $R_\text{NL}^\text{1f}$(Py) and $R_\text{NL}^\text{1f}$(Pt) follows from the rotation of $M_\text{Py}$. 
	At low $B$, $M_\text{Py}$ is aligned along the easy axis of the Py strips (y-axis, see definition of axes in Fig.~\ref{fig:DeviceGeometry}(c)), such that $\phi=0^\text{o}$ independently of $\theta$. In this regime, when $M_\text{Py}\parallel I$, there is no contribution from the ASHE. However, we still measure a finite amplitude of $R_\text{NL}^\text{1f}$(Py) and $R_\text{NL}^\text{1f}$(Pt), which we attribute to the magnons generated due to the SHE in Py, which is independent of $M_\text{Py}$ \cite{tian_manipulation_2016}. This contribution due to SHE, denoted as $R_\text{SHE}$ in Figs.~\ref{fig:ExperimentalData}(a) and \ref{fig:ExperimentalData}(b), remains approximately constant for low $B$. 
	As $B$ is further increased above 10~mT, $M_\text{Py}$ begins to tilt from the easy axis ($\phi \neq 0^\text{o}$), leading to a finite contribution towards magnon generation due to the ASHE. This contribution will be maximum when $M_\text{Py} \perp I$, $\textit{i.e.}$ $\phi = \pm 90^\text{o}$, which corresponds to $M_\text{Py}$ aligned along the hard axis of the Py strips (x-axis). The hard axis orientation of $M_\text{Py}$ is achieved for $B\approx 50$~mT, above which $R_\text{NL}^\text{1f}$(Py) and $R_\text{NL}^\text{1f}$(Pt) are saturated. Thus in this regime, both ASHE and SHE contribute, quantified as $R_\text{ASHE+SHE}$ in Figs.~\ref{fig:ExperimentalData}(a) and \ref{fig:ExperimentalData}(b).
	
	\begin{figure}[tbp]
		\includegraphics*[angle=0, trim=0mm 0mm 0mm 0mm, width=85mm]{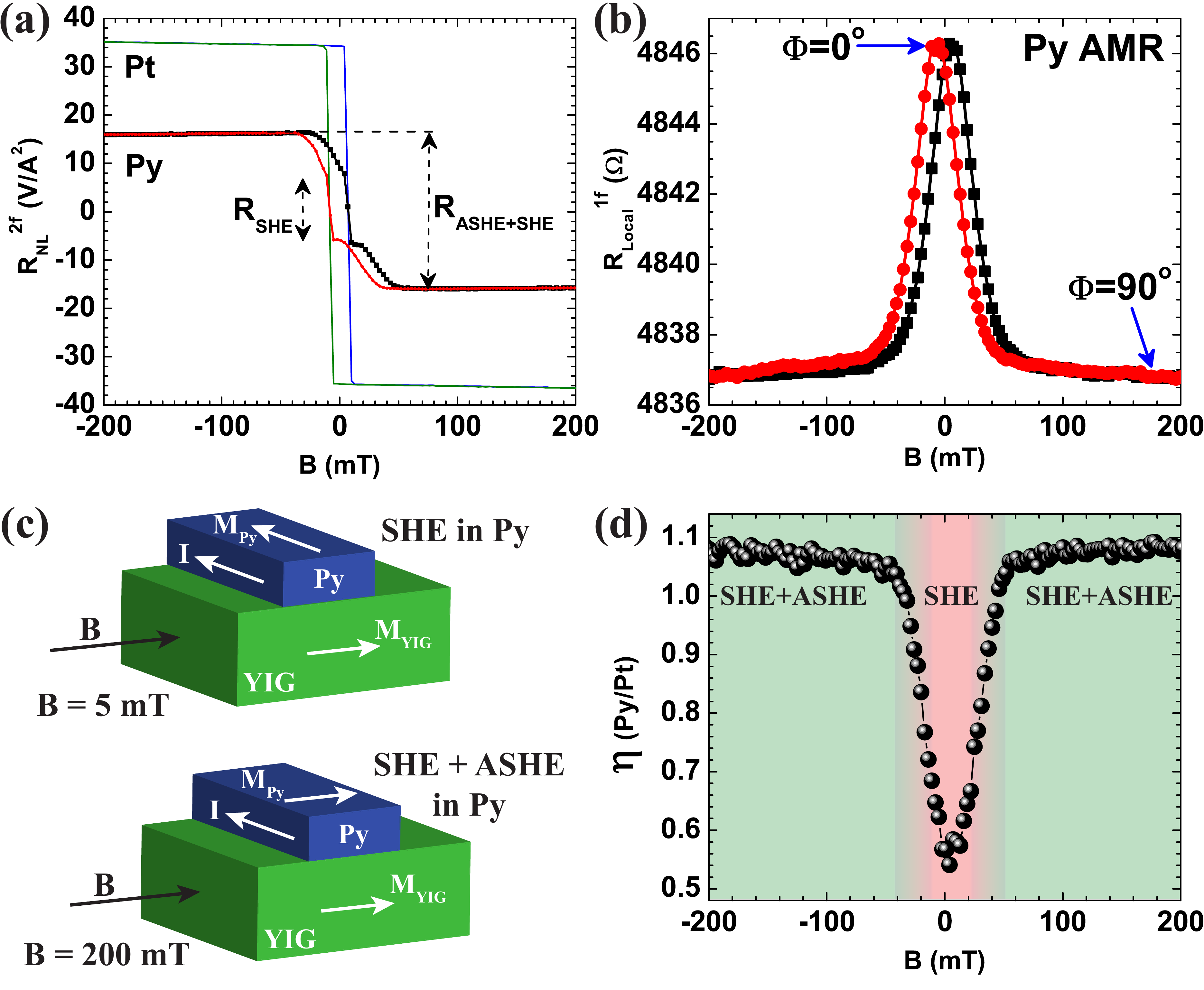}
		\caption{
			\label{fig:AMR&Efficiency}
			\textbf{(a)} The second harmonic response of the non-local resistance ($R_\text{NL}^\text{2f}$) as a function of $B$, for $\theta=90^\text{o}$. $R_\text{NL}^\text{2f}$ measured by both the Pt and the Py detectors shows a sharp switch around $B=0$, corresponding to the switching of $M_\text{YIG}$. The additional feature, only for the case of the Py detector, is due to the hard axis alignment of $M_\text{Py}$. \textbf{(b)} AMR measurement of the Py injector, exhibiting the saturation of $M_\text{Py}$ along the hard axis at $B\approx50$~mT. \textbf{(c)} Schematic representation of $M_\text{Py}$ with respect to $I$ for two different magnetic fields (5~mT and 200~mT). \textbf{(d)} The relative detection efficiency of Py over Pt ($\eta$(Py/Pt)), as a function of $B$, for $\theta=90^\text{o}$.
		}
	\end{figure}
	
	We also measure the second harmonic response $R_\text{NL}^\text{2f}$ for both the Py and Pt detectors, as well as the anisotropic resistance (AMR) of the Py strips, as shown in Figs.~\ref{fig:AMR&Efficiency}(a) and \ref{fig:AMR&Efficiency}(b), respectively. 
	The thermally generated magnons due to Joule heating at the Py injector produce the $R_\text{NL}^\text{2f}$ signal at the detector, via the spin Seebeck effect \cite{cornelissen_long-distance_2015}. Thus $R_\text{NL}^\text{2f}$ is independent of the magnetization of the injector. In Fig.~\ref{fig:AMR&Efficiency}(a), $R_\text{NL}^\text{2f}$ measured by the Pt detector exhibits a sharp switch around 0~mT, corresponding to the switching of $M_\text{YIG}$. A similar sharp switch is observed in the $R_\text{NL}^\text{2f}$ measured by the Py detector, only now it is followed by a gradual hard axis saturation of $M_\text{Py}$, up to $B\approx50$~mT. Thus from $R_\text{NL}^\text{2f}$(Py), we can clearly identify the separate behaviour of $M_\text{YIG}$ and $M_\text{Py}$, suggesting the lack of any strong coupling between the two. 
	The hard axis saturation of $M_\text{Py}$ is unambiguously confirmed from the AMR measurement presented in Fig.~\ref{fig:AMR&Efficiency}(b), in which the local resistance (2-probe) of the Py injector is measured as a function of $B$ for $\theta=90^\text{o}$. It clearly shows that $B\approx50$~mT is required to align $M_\text{Py} \perp I$, which corresponds accurately with the non-local data in Figs.~\ref{fig:ExperimentalData} and \ref{fig:AMR&Efficiency}(a). The orientations of $M_\text{Py}$ and $M_\text{YIG}$ with respect to $I$ in the Py injector, for two different magnetic field strengths, are illustrated in Fig.~\ref{fig:AMR&Efficiency}(c). These observations strongly support our hypothesis of two different contributions: ASHE and SHE.
	
	We now directly compare the magnon detection efficiencies of Py and Pt in the same device. Since the spin resistance of the medium (YIG) is much larger than the spin resistances of the injector and detectors \cite{cornelissen_magnon_2016}, the measured non-local resistance can be expressed as a product of the injection efficiency ($\eta_\text{I}$) of the injector and detection efficiency ($\eta_\text{D}$) of the detector. $\eta_\text{I}$ is the ratio of the spin accumulation created at the injector/YIG interface to the charge current sourced through the injector, whereas $\eta_\text{D}$ is the ratio of the measured non-local voltage in the detector to the spin current flowing across the YIG/detector interface. Thus, $R_\text{NL}^\text{1f}$(Py)~$\propto \eta_\text{I}$(Py)$\eta_\text{D}$(Py) and $R_\text{NL}^\text{1f}$(Pt)~$\propto \eta_\text{I}$(Py)$\eta_\text{D}$(Pt), since we use the same Py injector in both cases. The relative detection efficiency of Py to Pt can be then expressed as $\eta$(Py/Pt)~$=R_\text{NL}^\text{1f}$(Py)$/R_\text{NL}^\text{1f}$(Pt)~$= \eta_\text{D}$(Py)$/\eta_\text{D}$(Pt). In the lack of any theoretical study on ASHE, we phenomenologically express the dependence of the non-local resistance by updating Eq.~3 of Ref.~\cite{miao_inverse_2013}:
	
	\begin{equation}
	\eta_\text{D}\text{(Py)}\propto (\theta_\text{SH}^\text{Py}+\theta_\text{ASH}^\text{Py})\frac{\lambda_\text{Py}}{t_\text{Py}\sigma_\text{Py}}\tanh (\frac{t_\text{Py}}{2 \lambda_\text{Py}}),
	\label{eq:SpinHallAngle_NonlocalSignal}
	\end{equation}
	
	where, $\theta_\text{SH}^\text{Py}$ is the spin Hall angle in Py, $\theta_\text{ASH}^\text{Py}$ is the anomalous spin Hall angle, accounting for the spin-charge conversion in Py via the ASHE, $\lambda_\text{Py}$, $\sigma_\text{Py}$ and $t_\text{Py}$ being the spin relaxation length, electrical conductivity and the thickness of the Py strip, respectively. Considering $\lambda_\text{Py}=2.5$~nm \cite{miao_inverse_2013} and $t_\text{Py}=13$~nm, $\tanh (\frac{t_\text{Py}}{2 \lambda_\text{Py}})\approx1$. $\eta_\text{D}$(Pt) can be expressed similarly as relation~\ref{eq:SpinHallAngle_NonlocalSignal}, with the absence of the anomalous spin Hall angle in Pt. Considering $\lambda_\text{Pt}=1.5$~nm \cite{cornelissen_magnon_2016} and $t_\text{Pt}=7$~nm, $\tanh (\frac{t_\text{Pt}}{2 \lambda_\text{Pt}})\approx1$. 
	For accurately comparing the detection efficiencies of Py and Pt (considering that $\theta_\text{(A)SH}$, $\lambda$ and $\sigma$ are material specific properties), we account for the difference in their thicknesses and redefine $\eta$(Py/Pt)~$= [\eta_\text{D}\text{(Py)} \cdot t_\text{Py}]/[\eta_\text{D}\text{(Pt)} \cdot t_\text{Pt}]$. 
	In Fig.~\ref{fig:AMR&Efficiency}(d), $\eta$(Py/Pt) is plotted against $B$. The detection efficiency of Py exceeds that of Pt [($\eta$(Py/Pt)~$>1$)] in the SHE+ASHE regime, where the ASHE in Py is maximized. In the SHE only regime, the detection efficiency of Py is about 55\% that of Pt. 
	These observations show that the SHE and ASHE contributions in Py have the same polarity as the SHE in Pt. Note that since the electrical injection and detection are linear processes, the injection efficiency is equivalent to the detection efficiency. We therefore demonstrate an efficient and tunable magnon injection and detection process in Py by manipulating $M_\text{Py}$, switching on and off the contribution from the ASHE.
	
%
%

	\begin{figure}[tbp]
		\includegraphics*[angle=0, trim=0mm 0mm 0mm 0mm, width=85mm]{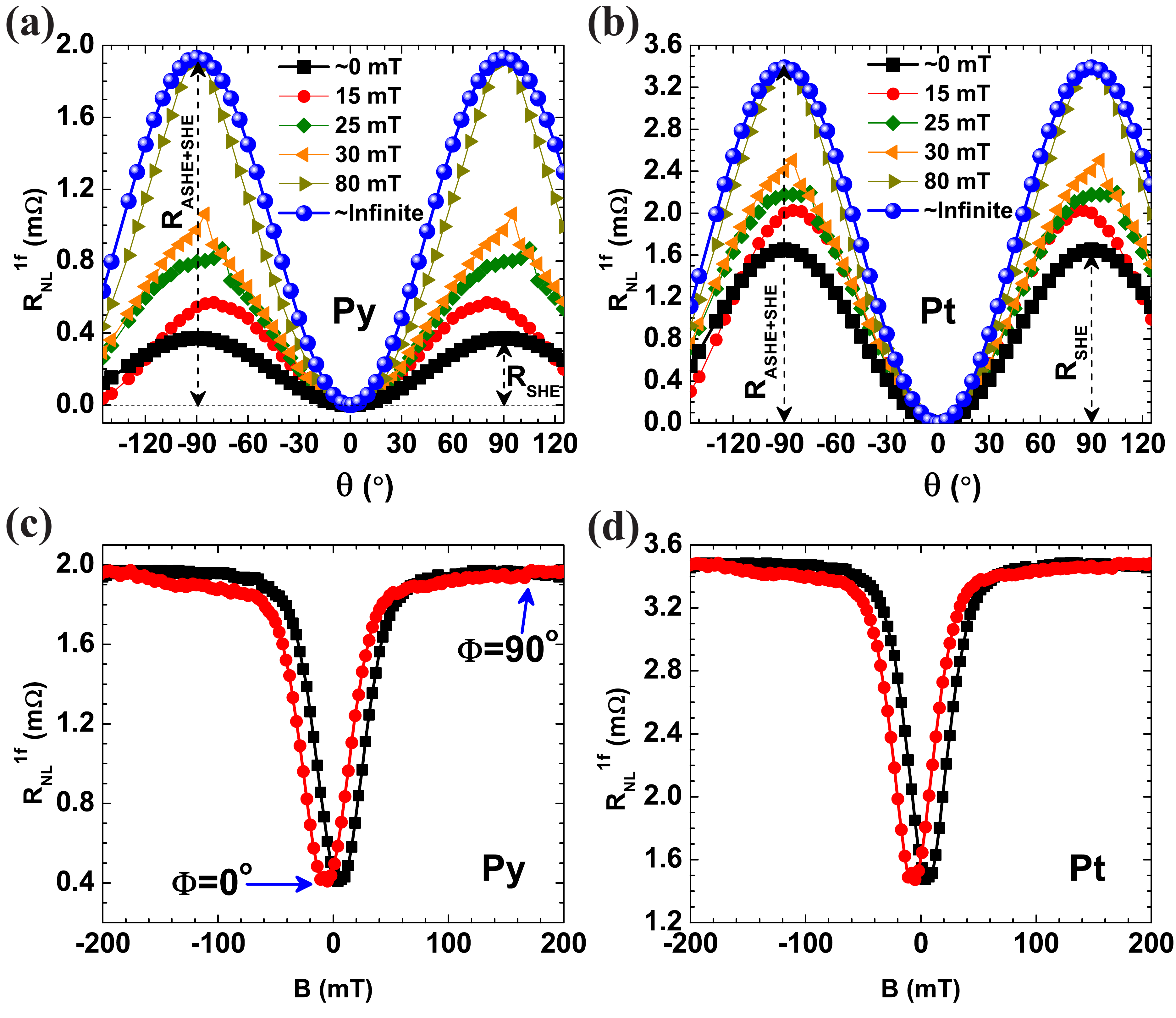}
		\caption{
			\label{fig:Simulated}
			 The modelled $R_\text{NL}^\text{1f}$(Py) and $R_\text{NL}^\text{1f}$(Pt) from Eqs.~\ref{eq:Py_detection} and \ref{eq:Pt_detection} are plotted against $\theta$ in \textbf{(a)} and \textbf{(b)}, respectively. The magnetic field dependence of $R_\text{NL}^\text{1f}$(Py) and $R_\text{NL}^\text{1f}$(Pt) is modelled in \textbf{(c)} and \textbf{(d)}, respectively. The simulated results exhibit an excellent agreement with the experimental data in Fig.~\ref{fig:ExperimentalData}. 
			 }
	\end{figure}

	We proceed to analytically formulate our hypothesis. The SHE will generate a spin accumulation in Py perpendicular to $I$, along the x-axis. The component of this spin accumulation parallel to $M_\text{YIG}$ will result in the generation of magnons in YIG. Thus the magnon generation due to the SHE will follow a  $\sin\theta$ dependence \cite{cornelissen_long-distance_2015} and will be independent of $M_\text{Py}$ \cite{tian_manipulation_2016}. On the other hand, the contribution due to the AHE is two-fold and proportional to $\sin\phi \cdot \cos(\theta-\phi)$. The first term $\sin\phi$ corresponds to the magnitude of the spin accumulation due to ASHE, controlled by the orthogonality between $I$ and $M_\text{Py}$, whereas the second term $\cos(\theta-\phi)$ corresponds to the projection of the spin accumulation due to ASHE (along $M_\text{Py}$) on $M_\text{YIG}$. The corresponding reciprocal processes will occur in the Py detector to generate $R_\text{NL}^\text{1f}$(Py). In the Pt detector, the spin to charge conversion will occur only via the ISHE and follow a $\sin\theta$ dependence. $R_\text{NL}^\text{1f}$(Py) and $R_\text{NL}^\text{1f}$(Pt) can therefore be expressed as:
	\begin{align}
	 R_\text{NL}^\text{1f}\text{(Py)}&=[a\sin\theta+b\sin\phi\cos(\theta-\phi)]^2, \label{eq:Py_detection}\\     
	R_\text{NL}^\text{1f}\text{(Pt)}&=c\sin\theta[a\sin\theta+b\sin\phi\cos(\theta-\phi)],
	\label{eq:Pt_detection}
	\end{align}
	where the coefficients $a$, $b$ and $c$ can be expressed as 
	$a\propto \frac{\theta_\text{SH}^\text{Py}\lambda_\text{Py}}{t_\text{Py}\sigma_\text{Py}}$, 
	$b\propto \frac{\theta_\text{ASH}^\text{Py}\lambda_\text{Py}}{t_\text{Py}\sigma_\text{Py}}$ and 
	$c\propto \frac{\theta_\text{SH}^\text{Pt}\lambda_\text{Pt}}{t_\text{Pt}\sigma_\text{Pt}}$. 
	Considering the case of $\phi=0^\text{o}$ and $\theta=90^\text{o}$ (low $B$) and equating Eq.~\ref{eq:Py_detection} to $R_\text{NL}^\text{1f}$(Py) obtained from Fig.~\ref{fig:ExperimentalData}(a), we calculate $a=0.61$~m$\Omega^{1/2}$. For $\phi=90^\text{o}$ and $\theta=90^\text{o}$ (high $B$), and substituting the value of $a$ in Eq.~\ref{eq:Py_detection}, we calculate $b=0.78$~m$\Omega^{1/2}$. Using these values of $a$ and $b$ and Eq.~\ref{eq:Pt_detection}, we find $c=2.58$~m$\Omega^{1/2}$. 
	Next, for simulating the angular dependence measurements, we first consider the two extreme cases: i) the high $B$ regime ($B\approx\infty$), where $M_\text{Py}$ is always aligned parallel to $M_\text{YIG}$, such that $\phi=\theta$ and ii) the low $B$ regime ($B\approx0$), where $M_\text{Py}$ is always aligned parallel to $I$, such that $\phi=0^\text{o}$. Substituting the values of the coefficients calculated above in Eqs.~\ref{eq:Py_detection} and \ref{eq:Pt_detection}, we model the angular dependence of $R_\text{NL}^\text{1f}$(Py) and $R_\text{NL}^\text{1f}$(Pt), as shown in Figs.~\ref{fig:Simulated}(a) and (b), respectively. For the intermediate regime of $B$ ($0<B<\infty$), we use the Stoner-Wohlfart model \cite{tannous_stoner-wohlfarth_2006} to calculate the dependence of $\phi$ on $\theta$ for different values of $B$, assuming a simple uniaxial shape anisotropy for $M_\text{Py}$, in order to simulate the angular dependence for different magnitudes of $B$. For modelling the $B$-sweep measurements, we extract the dependence of $\phi$ on $B$ from the AMR measurement in Fig.~\ref{fig:AMR&Efficiency}(b), following the expression \cite{rijks_semiclassical_1995,das_anisotropic_2016} $R_{\text{Py}}(B)=R_{\text{Py}}(\phi=90^\text{o})+[R_{\text{Py}}(\phi=0^\text{o})-R_{\text{Py}}(\phi=90^\text{o})]\cos^{2}\phi(B)$. The modelled results for the $B$-sweep measurements, using the same coefficients, are shown in Fig.~\ref{fig:Simulated}(c) and (d) for the Py and the Pt detectors, respectively. All the modelled results exhibit an excellent agreement with the experimental data both in terms of lineshapes and magnitudes of the non-local resistances. Finally, assuming similar spin mixing conductances between Py/YIG and Pt/YIG, we can approximately calculate the ratio $\theta_\text{SH}^\text{Py}/\theta_\text{SH}^\text{Pt} \approx (a\frac{t_\text{Py}\sigma_\text{Py}}{\lambda_\text{Py}})/(c\frac{t_\text{Pt}\sigma_\text{Pt}}{\lambda_\text{Pt}})=0.09$, which is lower to the value (0.38) reported by Miao \textit{et.al.}~\cite{miao_inverse_2013}. Additionally, we can estimate the ratio of the magnetization-dependent anomalous spin Hall angle to the magnetization-independent spin Hall angle in Py, $\theta_\text{ASH}^\text{Py}/\theta_\text{SH}^\text{Py} \approx b/a = 1.28$. What directly follows from this analysis is the ratio of the total charge-to-spin conversion in Py via the SHE and ASHE, to that in Pt only due to SHE, $\frac{\theta_\text{SH}^\text{Py}+\theta_\text{ASH}^\text{Py}}{\theta_\text{SH}^\text{Pt}}\approx 0.21$. 

	In this study, we have demonstrated a new spin injection and detection mechanism via the ASHE in Py, which can be tuned by an external magnetic field via manipulation of $M_\text{Py}$. We also found a finite contribution to the spin accumulation generated at the Py/YIG interface due to the SHE, independent of $M_\text{Py}$. This spin accumulation along the x-axis is non-trivial, since one would expect the spins to dephase under the influence of the exchange field of $M_\text{Py}$ which is oriented along the y-axis at low magnitudes of $B$. 
	Following a previous report of ISHE in Co being unaffected by its magnetization \cite{tian_manipulation_2016}, we conjecture that in Py (with lower magnetization) such dephasing is similarly negligible.
	Our work opens up the usage of ferromagnets as efficient and tunable sources of perpendicular spin current injection by electrical means. 
	By recognizing ASHE as a fundamental contributor to spin-to-charge conversion, we hope to inspire further investigations on implications of the coexistent spin-orbit driven phenomena in ferromagnets.

	\begin{acknowledgments}
		We acknowledge J.\ G.\ Holstein, H.\ M.\ de Roosz, H.\ Adema and T.\ Schouten for their technical assistance and thank G.\ E.\ W.\ Bauer, L.\ J.\ Cornelissen and J.\ Liu for discussions. We acknowledge the financial support of the Zernike Institute for Advanced Materials and	the Future and Emerging Technologies (FET) programme within the Seventh Framework Programme for Research of the European Commission, under FET-Open Grant No.~618083 (CNTQC).		
	\end{acknowledgments}

\bibliographystyle{apsrev4-1}

%

\end{document}